\documentclass[3p,twocolumn,longtitle,reversenotenum,sort&compress,lefttitle]{elsarticle}

\pdfoutput=1

\usepackage[utf8]{inputenc}
\usepackage[T1]{fontenc}

\PassOptionsToPackage{hyphens}{url}\usepackage[hidelinks]{hyperref}

\usepackage{libertine}
\usepackage{microtype}

\usepackage{enumitem,amssymb}
\usepackage{tabularx,booktabs}

\makeatletter
\let\@afterindenttrue\@afterindentfalse
\makeatother

\nopreprintlinetrue

\begin{document}

\begin{frontmatter}

\title{Recommendations on test datasets for evaluating AI
solutions in pathology}

\author[aff1]{André~Homeyer\fnref{equalcontr}\corref{corresp}}
\ead{andre.homeyer@mevis.fraunhofer.de}
\author[aff2]{Christian~Geißler\fnref{equalcontr}}
\author[aff1]{Lars~Ole~Schwen\fnref{equalcontr}}
\author[aff3]{Falk~Zakrzewski\fnref{equalcontr}}
\author[aff2]{Theodore~Evans\fnref{equalcontr}}
\author[aff4]{Klaus~Strohmenger\fnref{equalcontr}}
\author[aff1]{Max~Westphal\fnref{equalcontr}}
\author[aff5]{Roman~David~Bülow\fnref{equalcontr}}
\author[aff6]{Michaela~Kargl}
\author[aff2]{Aray~Karjauv}
\author[aff7]{Isidre~Munné-Bertran}
\author[aff2]{Carl~Orge~Retzlaff}
\author[aff8]{Adrià~Romero-López}
\author[aff9]{Tomasz~Sołtysiński}
\author[aff6]{Markus~Plass}
\author[aff4]{Rita~Carvalho}
\author[aff10]{Peter~Steinbach}
\author[aff5]{Yu-Chia~Lan}
\author[aff5]{Nassim~Bouteldja}
\author[aff8]{David~Haber}
\author[aff8]{Mateo~Rojas-Carulla}
\author[aff5]{Alireza~Vafaei~Sadr}
\author[aff8]{Matthias~Kraft}
\author[aff11]{Daniel~Krüger}
\author[aff12]{Rutger~Fick}
\author[aff13]{Tobias~Lang}
\author[aff5]{Peter~Boor}
\author[aff6]{Heimo~Müller}
\author[aff4]{Peter~Hufnagl}
\author[aff4]{Norman~Zerbe}

\address[aff1]{Fraunhofer Institute for Digital Medicine MEVIS, Max-von-Laue-Straße 2, 28359 Bremen, Germany}
\address[aff2]{Technische Universität Berlin, DAI-Labor, Ernst-Reuter-Platz 7, 10587 Berlin, Germany}
\address[aff3]{Institute of Pathology, Carl Gustav Carus University Hospital Dresden (UKD), TU Dresden (TUD), Fetscherstrasse 74, 01307 Dresden, Germany}
\address[aff4]{Charité -- Universitätsmedizin Berlin, corporate member of Freie Universität Berlin and Humboldt Universität zu Berlin, Institute of Pathology, Charitéplatz 1, 10117 Berlin, Germany}
\address[aff5]{Institute of Pathology, University Hospital RWTH Aachen, Pauwelsstraße 30, 52074 Aachen, Germany}
\address[aff6]{Medical University of Graz, Diagnostic and Research Center for Molecular BioMedicine, Diagnostic \& Research Institute of Pathology, Neue~Stiftingtalstrasse~6, 8010 Graz, Austria}
\address[aff7]{MoticEurope, S.L.U., C. Les Corts, 12 Poligono Industrial, 08349 Barcelona, Spain}
\address[aff8]{Lakera AI AG, Zelgstrasse 7, 8003 Zürich, Switzerland}
\address[aff9]{QuIP GmbH, Reinhardtstraße 1, 10117 Berlin, Germany}
\address[aff10]{Helmholtz-Zentrum Dresden Rossendorf, Bautzner Landstraße 400, 01328 Dresden, Germany}
\address[aff11]{Olympus Soft Imaging Solutions GmbH, Johann-Krane-Weg 39, 48149 Münster, Germany}
\address[aff12]{Tribun Health, 2 Rue du Capitaine Scott, 75015 Paris, France}
\address[aff13]{Mindpeak GmbH, Zirkusweg 2, 20359 Hamburg, Germany}

\fntext[equalcontr]{These authors contributed equally to this work.}
\cortext[corresp]{Corresponding author}

\begin{abstract}
Artificial intelligence (AI) solutions that automatically extract information from digital histology images have shown great promise for improving pathological diagnosis. Prior to routine use, it is important to evaluate their predictive performance and obtain regulatory approval. This assessment requires appropriate test datasets. However, compiling such datasets is challenging and specific recommendations are missing.

A committee of various stakeholders, including commercial AI developers, pathologists, and researchers, discussed key aspects and conducted extensive literature reviews on test datasets in pathology. Here, we summarize the results and derive general recommendations for the collection of test datasets.

We address several questions: Which and how many images are needed? How to deal with low-prevalence subsets? How can potential bias be detected? How should datasets be reported? What are the regulatory requirements in different countries?

The recommendations are intended to help AI developers demonstrate the utility of their products and to help regulatory agencies and end users verify reported performance measures. Further research is needed to formulate criteria for sufficiently representative test datasets so that AI solutions can operate with less user intervention and better support diagnostic workflows in the future.
\end{abstract}

\end{frontmatter}

\section{Introduction}

The application of artificial intelligence techniques to digital tissue images has shown great promise for improving pathological diagnosis~\citep{serag2019,abels2019,moxley-wiles2020}. They can not only automate time-consuming diagnostic tasks and make analyses more sensitive and reproducible, but also extract new digital biomarkers from tissue morphology for precision medicine~\citep{echle2021}.

Pathology involves a large number of diagnostic tasks, each being a potential application for AI. Many of these involve the characterization of tissue morphology. Such tissue classification approaches have been developed for identifying tumors in a variety of tissues, including lung~\citep{coudray2018,wang2020}, colon~\citep{iizuka2020}, breast~\citep{cruz-roa2017,campanella2019}, and prostate~\citep{campanella2019} but also in non-tumor pathology, e.g., kidney transplants~\citep{kers2022}. Further applications include predicting outcomes~\citep{skrede2020,saillard2020} or gene mutations~\citep{kather2019,coudray2018,couture2018} directly from tissue images. Similar approaches are also employed to detect and classify cell nuclei, e.g., to quantify the positivity of immunohistochemistry markers like Ki67, ER/PR, Her2, and PD-L1~\citep{hoefener2018,balkenhol2021}.

Testing AI solutions is an important step to ensure that they work reliably and robustly on routine laboratory cases. AI algorithms run the risk of exploiting feature associations that are specific to their training data~\citep{lever2016}. Such ``overfitted'' models tend to perform poorly on previously unseen data. To obtain a realistic estimate of the prediction performance on real-word data, it is common practice to apply AI solutions to a test dataset. The results are then compared with reference results in terms of task-specific performance metrics, e.g., sensitivity, specificity, or ROC-AUC.

Test datasets may only be used once to evaluate the performance of a finalized AI solution~\citep{lever2016}. They may not be considered during development. This can be considered a consequence of Goodhart's law stating that measures cease to be meaningful when used as targets~\citep{strathern1997}: If AI solutions are optimized for test datasets, they cannot provide realistic performance estimates for real-world data. Test datasets are also referred to as ``hold-out datasets'' or ``(external) validation datasets.'' The term ``validation,'' however, is not used consistently in the machine learning community and can also refer to model selection during development~\citep{lever2016}.

Besides overfitting, AI methods are prone to ``shortcut learning''~\citep{geirhos2020}. Many datasets used in the development of AI methods contain confounding variables (e.g., slide origin, scanner type, patient age) that are spuriously correlated with the target variable (e.g., tumor type)~\citep{schmitt2021}. AI methods often exploit features that are discriminative for such confounding variables and not for the target variable~\citep{wallis2022}. Despite working well for smaller datasets containing similar correlations, such methods fail in more challenging real-world scenarios in ways humans never would~\citep{oakden-rayner2020}. To minimize the likelihood of spurious correlations between confounding variables and the target variable, test datasets must be large and diversified~\citep{schmitt2021}. At the same time, test datasets must be small enough to be acquired with realistic effort and cost. Finding a good balance between these requirements is a major challenge for AI developers.

Comparatively little attention has been paid to compiling test datasets for AI solutions in pathology. Datasets for training, on the other hand, were considered frequently~\citep{campanella2019,nagpal2019,tang2021,vali-betts2021,tellez2019,anghel2019,maree2017}. Training datasets are collected with a different goal than test datasets: While training datasets should produce the best possible AI models, test datasets should provide the most realistic performance assessment for routine use, which presents unique challenges.

Some publications address individual problems in compiling test datasets in pathology, e.g., how to avoid bias in the performance evaluation caused by site-specific image features in test datasets~\citep{howard2021}. Other publications provide general recommendations for evaluating AI methods for medical applications without considering the specific challenges of pathology~\citep{oala2020,maleki2020,cabitza2021,park2021,dehond2022}.

Appropriate test datasets are critical to demonstrate the utility of AI solutions as well as to obtain regulatory approval. However, the lack of guidance on how to compile test datasets is a major barrier to the adoption of AI solutions in laboratory practice.

This article gives recommendations for test datasets in pathology. It summarizes the results of extensive literature reviews and discussions by a committee of various stakeholders, including commercial AI developers, pathologists, and researchers. This committee was established as part of the EMPAIA project (Ecosystem for Pathology Diagnostics with AI Assistance), aiming to facilitate the adoption of AI in pathology~\citep{hufnagl2021}.

\section{Results}

The next sections discuss and provide recommendations on various aspects that must be considered when creating test datasets. For meaningful performance estimates, test datasets must be both diverse enough to cover the variability of data in routine diagnostics and large enough to allow statistically meaningful analyses. Relevant subgroups must be covered, and test datasets should be unbiased. Moreover, test datasets must be sufficiently independent of datasets used in the development of AI solutions. Comprehensive information about test datasets must be reported and regulatory requirements must be met when evaluating the clinical applicability of AI solutions.

\subsection{Target population of images}

All images an AI solution may encounter in its intended use constitute its ``target population of images.'' A test dataset must be an adequate sample of this target population to provide a reasonable estimate of the prediction performance of the AI solution. For all applications in pathology, the target population is distributed across multiple dimensions of variability, see~\autoref{tab:data_variabilities}.

\begin{table*}[t!]
\centering\small
\newenvironment{compact_itemize}%
    {\begin{minipage}[t]{\linewidth}\raggedright\begin{itemize}[leftmargin=2em,nosep]}%
    {\end{itemize}\end{minipage}}
\begin{tabularx}{\textwidth}{l@{\qquad}X}
\toprule
\textbf{Origin} & \textbf{Variabilities} \\
\midrule
Patient &
\begin{compact_itemize}
    \item Patient ethnicity
    \item Patient demographics
    \item Disease stage/severity
    \item Rare cases of disease
    \item Comorbidities
    \item Biological differences (genetic, transcriptional, epigenetic, proteomic, and metabolomic)
\end{compact_itemize}\\
\midrule
Specimen sampling &
\begin{compact_itemize}
    \item Tissue heterogeneity
    \item Size of tissue section
    \item Coverage of diseased/healthy/boundary regions
    \item Tissue damage, e.g., torn, cauterized
    \item Surgical ink present
\end{compact_itemize}\\
\midrule
Slide processing &
\begin{compact_itemize}
    \item Inter-material and device differences
    \item Preparation differences (fixation, dehydration; freezing; mechanical handling)
    \item Cutting artifacts (torn, folded, deformed, thick or inhomogeneously thick tissue)
    \item Foreign matter/floaters in specimen
    \item Over-/under-staining, inhomogeneous staining
    \item Foreign objects on slide/cover slip (dirt, stain residue, pen markings, fingerprint)
    \item Cracks, air bubbles, scratches
    \item Slide age
\end{compact_itemize}\\
\midrule
Imaging/image processing &
\begin{compact_itemize}
    \item Inter- and intra-scanner differences
    \item Out-of-focus images, heterogeneous focus
    \item Amount of background in analyzed image region
    \item Magnification/image resolution
    \item Heterogeneous illumination
    \item Grid noise, stitching artifacts
    \item Lossy image compression
\end{compact_itemize}\\
\midrule
Ground truth annotation &
\begin{compact_itemize}
    \item Inter- and intra-observer differences
    \item Ambiguous cases
\end{compact_itemize}\\
\bottomrule
\end{tabularx}
\caption{Examples of data variabilities within the intended use~\citep{chen2021,avanaki2016,schoemig-markiefka2021,focke2017,tellez2019,taqi2018,chatterjee2014,schmitt2021,cajal2020,pursnani2016}.}
\label{tab:data_variabilities}
\end{table*}

\begin{figure*}
\centering
\includegraphics[width=.8\textwidth]{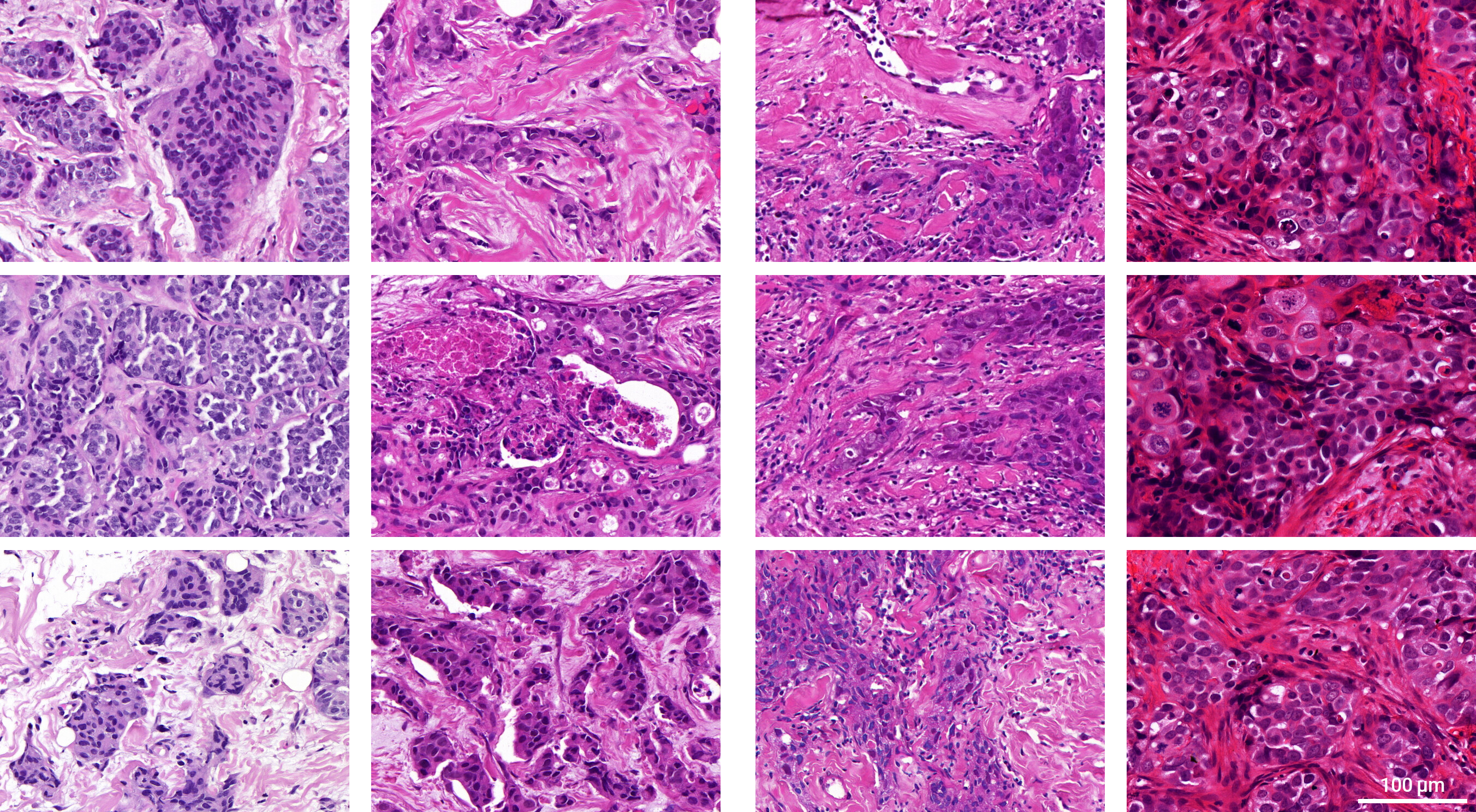}

\caption{Examples of tissue variability within and between biopsies (H\&E-stained breast tissue of female patients with invasive carcinomas of no special type, 40× objective magnification). First and second column from the left: 41yo patients, grade 2; third and fourth column: 42yo patients, grade 3.}
\label{fig:variability}
\end{figure*}

\textit{Biological variability.} The visual appearance of tissue varies between normal and diseased states. This is what AI solutions are designed to detect and characterize. But even tissue of the same category can look very different (see \autoref{fig:variability}). The appearance is influenced by many factors (e.g., genetic, transcriptional, epigenetic, proteomic, and metabolomic) that differ between patients as well as between demographic and ethnic groups~\citep{cajal2020}. These factors often vary spatially (e.g., different parts of organs are differently affected) and temporally (e.g., the pathological alterations differ based on disease stage) within a single patient~\citep{dagogo-jack2017}.

\textit{Technical variability.} Processing and digitization of tissue sections consists of several steps (e.g., tissue fixation, processing, cutting, staining and digitization) all of which can contribute to image variability~\citep{chen2021}. Differences in section thickness and staining solutions can lead to variable staining appearances ~\citep{focke2017}. Artifacts frequently occur during tissue processing, including elastic deformations, inclusion of foreign objects, and cover glass scratches~\citep{schoemig-markiefka2021}. Differences in illumination, resolution, and encoding algorithms of slide scanner models also affect the appearance of tissue images~\citep{chen2021}.

\textit{Observer variability.} Images in test datasets are commonly associated with a reference label like a disease category or score determined by a human observer. It is well known that the assessment of tissue images is subject to intra- and inter-observer variability~\citep{allison2014,el-badry2009,martinez2007,kujan2007,boiesen2000,oni2017,furness2003}. This variability results from subjective biases (e.g., caused by training, specialization, and experience) but also from inherent ambiguities in the images~\citep{tizhoosh2021,homeyer2017}.

\begin{figure*}[t]
\centering
\includegraphics[width=.7\textwidth]{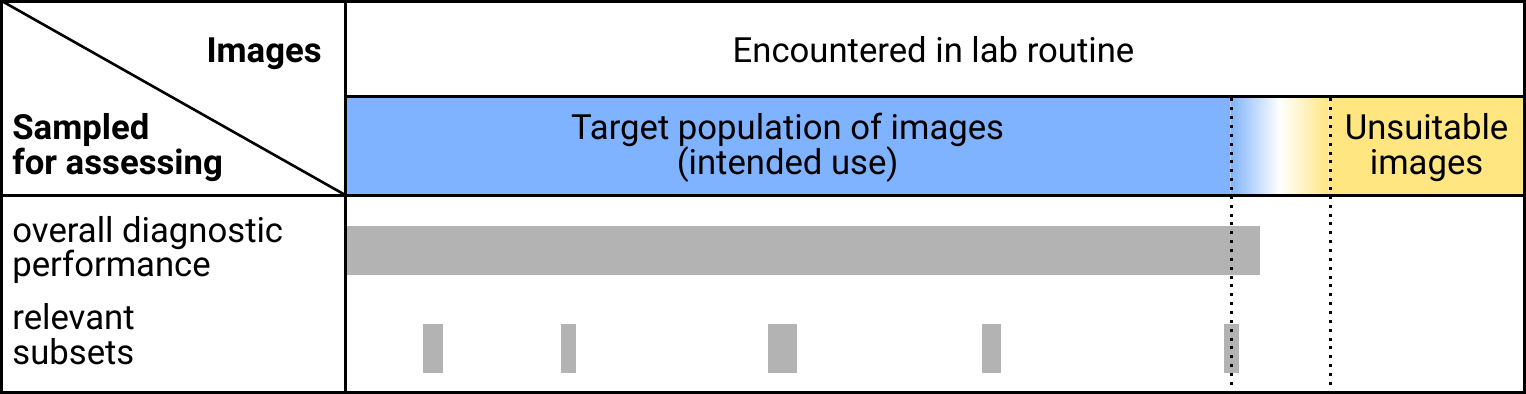}

\caption{Qualitative overview of sampling regimes for performance assessment in the entire target population of images or in specific subgroups. The boundary between the target population of images and unsuitable images that do not fall under the intended use is fuzzy.}
\label{fig:sampling_regimes}
\end{figure*}

\begin{figure*}[t]
\centering
\includegraphics[width=.9\textwidth]{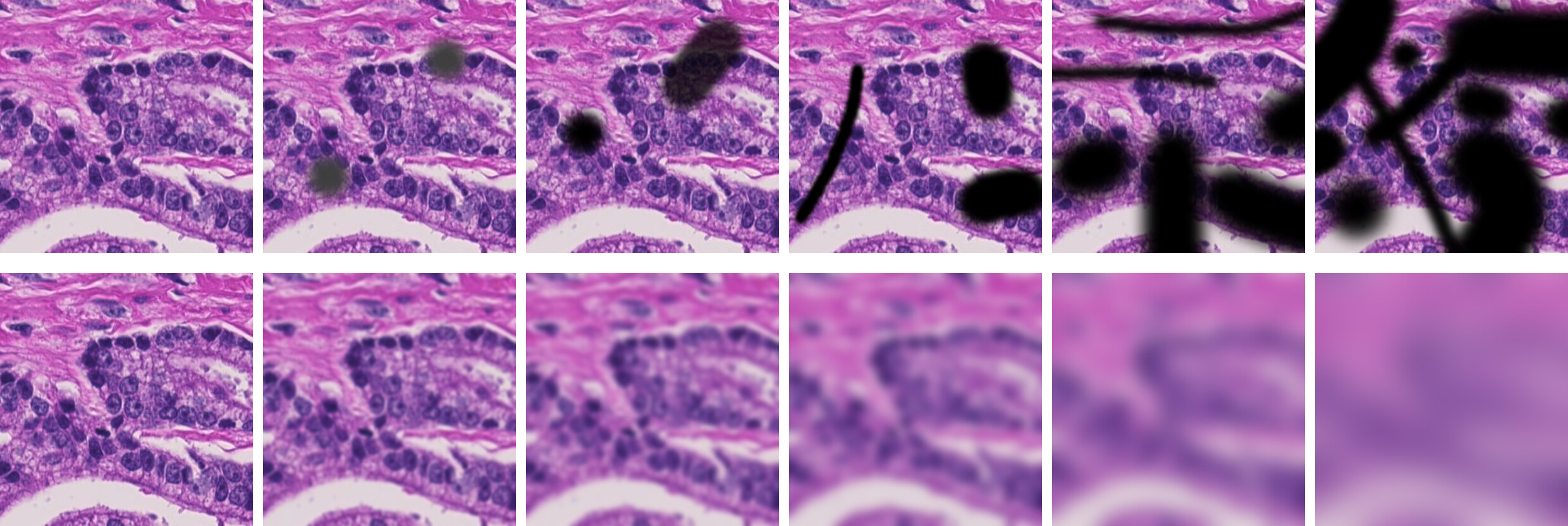}

\caption{Examples of different severity levels of artifacts on a prostate section. The top row shows simulated foreign objects, the bottom row shows simulated focal blur. The original image on the left is clearly within the intended use of algorithms for Gleason grading in prostate cancer diagnostics, while the rightmost images are clearly unsuitable. The tissue image is adapted from another source~\citep{arvaniti2018dataset} (CC0-licensed~\citep{CC0license}).}
\label{fig:severity_levels}
\end{figure*}

Routine laboratory work occasionally produces images that are unsuitable for the intended use of an AI solution, e.g., because they are ambiguous or of insufficient quality. Most AI solutions require prior quality assurance steps to ensure that solutions are only applied to suitable images~\citep{perincheri2021,dasilva2021}. The boundary between suitable and unsuitable images is usually fuzzy (see \autoref{fig:sampling_regimes}) and there are difficult images that cannot be clearly assigned to either category (see \autoref{fig:severity_levels}).

Defining the target population is challenging and presumes a clear definition of the intended use by the AI developer. The target population of images must be defined before test datasets are collected. It must be clearly stated which subsets of images fall under the intended use. Such subsets may consist of specific disease variants, demographic characteristics, ethnicities, staining characteristics, artifacts, or scanner types. These subsets typically overlap, e.g., the subset of images of one scanner type contains images from different patient age groups. A particular challenge is to define where the target population ends. Examples of images within and outside the intended use can help human observers sort out unsuitable images as objectively as possible.

\subsection{Data collection}

Test datasets must be representative of the entire target population of images, i.e., sufficiently diverse and unbiased. To minimize spurious correlations between confounding variables and the target variable and to uncover shortcut learning in AI methods, all dimensions of biological and technical variability must be adequately covered for the classes considered~\citep{schmitt2021,maree2017}, also reflecting the variability of negative cases without visible pathology~\citep{maree2017,ianni2020}.

All images encountered in the normal laboratory workflow must be considered. One way to achieve this is to collect all cases that occurred over a given time period~\citep{ianni2020} long enough for a sufficient number of cases to be collected (e.g., one year~\citep{campanella2019}). Data should be collected from multiple international laboratories, since they differ in their spectra of patients and diseases, technical equipment and operating procedures. To avoid selection bias, artifacts or atypical morphologies must not be excluded if they are part of the intended use of the product~\citep{campanella2019,ianni2020,freeman2021}. Data should be collected at the point in the workflow where the AI solution would be applied, taking into account possible prior quality assurance steps in the workflow.

All data in a test dataset must be collected according to a consistent acquisition protocol (see ``Reporting''). The best way to ensure this is to prospectively collect test datasets according to this protocol. Retrospective datasets were typically collected for a different purpose and are thus likely to be subject to selection bias, that is difficult to adjust for~\citep{talari2020}. If retrospective data are used in a test dataset, a comprehensive description of the acquisition protocol must be available so that potential issues can be identified~\citep{gianfrancesco2018}.

\subsubsection{Annotation}

Test datasets for AI solutions contain not only images, but also annotations representing the expected analysis result, e.g., slide-level labels or delineations of tissue regions. In most cases, such reference annotations must be prepared by human observers with sufficient experience in the diagnostic use case. Since humans are prone to intra- and inter-observer variability, annotations in test datasets should be created by multiple observers from different hospitals or laboratories. For unequivocal results, it can be helpful to organize consensus conferences and to use standardized electronic reporting formats~\citep{allison2014}. Any remaining disagreement should be documented with justification (e.g., suboptimal sample quality) and considered when evaluating AI solutions. Semi-automatic annotation methods can help reduce the effort required for manual annotation~\citep{gamper2020,graham2021}. However, they can introduce biases themselves and should therefore be monitored by human observers.

\subsubsection{Curation}

Unsuitable data that does not fit the intended use of an AI solution should not be included in a test dataset. Such data usually must be detected by human observers, e.g., in a dedicated data curation step or during the generation of reference annotations. However, there are automated tools to support this process~\citep{janowczyk2019}. Some approaches identify unsuitable data based on basic image features such as brightness, predominant colors, and sharpness~\citep{ameisen2014,senaras2018} or by detecting typical artifacts like tissue folds and air bubbles~\citep{avanaki2016,smit2021}. Other methods analyze domain shifts~\citep{stacke2021,bozorgtabar2021,linmans2020} or use dedicated neural networks trained for outlier detection~\citep{guha-roy2022}. There are also approaches for detecting outliers depending on the tested AI solution~\citep{calli2019,cao2020,berger2021,stacke2021,zhang2021}. Although these approaches can help exclude unsuitable images from test datasets, they do not yet appear to be mature enough to be used entirely without human supervision.

\subsubsection{Synthetic data}

There are a variety of techniques for extending datasets with synthetic data. Some techniques alter existing images in a generic (e.g., rotation, mirroring) or histology-specific way (e.g., stain transformations~\citep{tellez2019} or emulation of image artifacts~\citep{wang2021,sinha2021,schoemig-markiefka2021,lehmussola2007,ulman2016,gadermayr2019,moghadam2022}). Other techniques create fully synthetic images from scratch~\citep{niazi2018,levine2020,quiros2019,jose2021,deshpande2022}. These techniques are useful for data augmentation~\citep{janowczyk2016,serag2019,abels2019}, i.e., enriching development data in order to avoid overfitting and increase robustness. However, they cannot replace original real-world data for test datasets. Because all of these techniques are based on simplified models of real-world variability, they are likely to introduce biases into a test dataset and make meaningful performance measurement impossible.

\subsection{Sample size}

Any test dataset is a sample from the target population of images, thus any performance metric computed on a test dataset is subject to sampling error. In order to draw reliable conclusions from evaluation results, the sampling error must be sufficiently small. Larger samples generally result in lower sampling error, but are also more expensive to produce. Therefore, the minimum sample size required to achieve a maximum allowable sampling error should be determined prior to data collection.

Many different methods have been proposed for sample size determination. Most of them refer to statistical significance tests which are used to test a prespecified hypothesis about a population parameter (e.g., sensitivity, specificity, ROC-AUC) on the basis of an observed data sample~\citep{adcock1997,pepe2004,flahault2005}. Such sample size determination methods are commonly used in clinical trial planning and available in many statistical software packages~\citep{zhang2021}.

When evaluating AI solutions in pathology, the goal is more often to estimate a performance metric with a sufficient degree of precision than to test a previously defined hypothesis. Confidence intervals (CIs) are a natural way to express the precision of an estimated metric and should be reported instead of or in addition to test results~\citep{bland2009}. A CI is an interval around the sample statistic that is likely to cover the true population value at some confidence level, usually 95\%~\citep{hazra2017}. The sample statistic can either be the performance metric itself or a difference between the performance metrics of two methods, e.g., when comparing performance to an established solution.

When using CIs, the sample size calculation can be based on the targeted width of the CI which is inversely proportional to the precision of the performance estimation~\citep{bland2009}. Several approaches have been proposed for that matter~\citep{hanley1982,simel1991,kelley2003,riley2021,pavlou2021}. To determine a minimum sample size, assumptions regarding the sample statistic, its variability, and usually also its distributional form must be made. The open-source software ``presize'' implements several of these methods and provides a simple web-based user interface to perform CI-based sample size calculations for common performance metrics~\citep{haynes2021}.

\subsection{Subsets}

AI solutions that are very accurate on average often perform much worse on certain subsets of their target population of images~\citep{echle2022}, a phenomenon known as ``hidden stratification.'' Such differences in performance can exceed 20\%~\citep{oakden-rayner2020}. Hidden stratification occurs particularly in low-prevalence subgroups, but may also occur in subgroups with poor label quality or subtle distinguishing characteristics~\citep{oakden-rayner2020}. There are substantial differences in cancer incidence, e.g., by gender, socioeconomic status, and geographic region~\citep{sung2021}. Hence, hidden stratification may result in disproportionate harm to patients in less common demographic groups and jeopardize the clinical applicability of AI solutions~\citep{oakden-rayner2020}. Common performance measures computed on the entire test dataset can be dominated by larger subsets and do not indicate whether there are subsets for which an AI solution underperforms~\citep{saito2015}.

To detect hidden stratification, AI solutions must be evaluated independently on relevant subsets of the target population of images (e.g., certain medical characteristics, patient demographics, ethnicities, scanning equipment)~\citep{oakden-rayner2020,echle2022}. This means in particular that the metadata for identifying the subsets must be available~\citep{oala2020}. Performance evaluation on subsets is an important requirement to obtain clinical approval by the FDA (see ``Regulatory requirements''). Accordingly, such subsets should be specifically delineated within test datasets. Each subset needs to be sufficiently large to allow statistically meaningful results (see ``Sample size''). It is important to provide information on why and how subsets were collected so that any issues AI solutions may have with specific subsets can be specifically tracked (see ``Reporting''). Identifying subsets at risk of hidden stratification is a major challenge and requires extensive knowledge of the use case and the distribution of possible input images~\citep{oakden-rayner2020}. As an aid, potentially relevant subsets can also be detected automatically using unsupervised clustering approaches such as k-means~\citep{oakden-rayner2020}. If a detected cluster underperforms compared to the entire dataset, this may indicate the presence of hidden stratification that needs further examination.

\subsection{Bias detection}

Biases can make test datasets unsuitable for evaluating the performance of AI algorithms. Therefore, it is important to identify potential biases and to mitigate them early during data acquisition~\citep{maree2017}. Bias, in this context, refers to sampling bias, i.e., the test dataset is not a randomly drawn sample from the target population of images. Subsets to be evaluated independently may be biased by construction with respect to particular features (e.g., patient age). Here, it is important to ensure that the subgroups do not contain unexpected biases with respect to other features. For example, the prevalence of slide scanners should be independent of patient age, whereas the prevalence of diagnoses may vary by age group.

For features represented as metadata (e.g., patient age, slide scanner, or diagnosis), bias can be detected by comparing the feature distributions in the test dataset and the target population using summary statistics (e.g., via mean and standard deviation) or dedicated fairness metrics~\citep{qi2021,cabitza2020}. Detection of bias in an entire test dataset requires a good estimate of the feature distribution of the target population of images. Bias in subgroups can be detected by comparing the subset distribution to the entire dataset. Several toolkits for measuring bias based on metadata have been proposed~\citep{saleiro2018,bellamy2018} and evaluated~\citep{lee2021}.

Detecting bias in the image data itself is more challenging. Numerous features can be extracted from image data and it is difficult to determine the distribution of these features in the target population of images. Similar to automatic detection of unsuitable data, there are automatic methods to reveal bias in image data. Domain shifts~\citep{stacke2021} can be detected either by comparing the distributions of basic image features (e.g., contrast) or by more complex image representations learned through specific neural network models~\citep{stacke2021,guha-roy2022,roohi2020}. Another approach is to train trivial machine learning models with modified images from which obvious predictive information has been removed (e.g., tumor regions): If such models perform better than chance, this indicates bias in the dataset~\citep{model2015,shamir2008}.

\subsection{Independence}

In the development of AI solutions, it is common practice to split a given dataset into two sets, one for development (e.g., a training and a validation set for model selection) and one for testing~\citep{lever2016}. AI methods are prone to exploit spurious correlations in datasets as shortcut opportunities~\citep{geirhos2020}. In this case, the methods perform well on data with similar correlations, but not on the target population. If both development and test datasets are drawn from the same original dataset, they are likely to share spurious correlations, and the performance on the test dataset may overestimate the performance on the target population. Therefore, datasets used for development and testing need to be sufficiently independent. As explained below, it is not sufficient for test datasets to merely contain different images than development datasets~\citep{lever2016,geirhos2020}.

To account for memory constraints, histologic whole-slide images (WSIs) are usually divided into small sub-images called ``tiles.'' AI methods are then applied to each tile individually, and the result for the entire WSI is obtained by aggregating the results of the individual tiles. If tiles are randomly assigned, tiles from the same WSI can end up in both the development and the test datasets, possibly inflating performance results. A substantial number of published research studies are affected by this problem~\citep{bussola2021}. Therefore, to avoid any risk of bias, none of the tiles in a test dataset may originate from the same WSI as the tiles in the development set~\citep{bussola2021}.

Datasets can contain site-specific feature distributions~\citep{howard2021}. If these site-specific features are correlated with the outcome of interest, AI methods might use these features for classification rather than the relevant biological features (e.g., tissue morphology) and be unable to generalize to new datasets. A comprehensive evaluation based on multi-site datasets from TCGA showed that including data from one site in development and test datasets often leads to overoptimistic estimates of model accuracy~\citep{howard2021}. This study also found that commonly used color normalization and augmentation methods did not prevent models from learning site-specific features, although stain differences between laboratories appeared to be a primary source of site-specific features. Therefore, the images in development and test datasets must originate not only from different subjects, but should also from different clinical sites~\citep{maleki2020,wu2021,koenig2007}.

As described in the Introduction section, a given AI solution should only be evaluated once against a given test dataset~\citep{lever2016}. Datasets published in the context of challenges or studies (many of which are based on TCGA~\citep{echle2021} and have regional biases~\citep{celi2022}) should generally not be used as test datasets: it cannot be ruled out that they were taken into account in some form during development, e.g., inadvertently or as part of pretraining. Ideally, test datasets should not be published at all and the evaluation should be conducted by an independent body with no conflicts of interest~\citep{oala2020}.

\subsection{Reporting}

Adequate reporting of test datasets is essential to determine whether a particular dataset is appropriate for a particular AI solution. Detailed metadata on the coverage of various dimensions of variability is required for detecting bias and identifying relevant subsets. Data provenance must be tracked to ensure that test data are sufficiently disjoint from development data~\citep{maree2017,howard2021}. Requirements for the test data~\citep{ai4h2020del5-4} and acquisition protocols~\citep{ai4h2020del5-1} should also be reported so that further data can be collected later. Accurate reporting of test datasets is important in order to submit evaluation results traceable to the test data for regulatory approval~\citep{MDCG2022-2}.

Various guidelines for reporting clinical research and trials, including diagnostic models, have been published~\citep{moons2015}. Some of these have been adapted specifically for machine learning approaches~\citep{liu2020,norgeot2020} or such adaptation is under development~\citep{wiegand2019,wenzel2020,sounderajah2021,collins2021}. However, only very few guidelines elaborate on data reporting~\citep{stevens2020}, and there is not yet consensus on structured reporting of test datasets, particularly for computational pathology.

Data acquisition protocols should comprehensively describe how and where the test dataset was acquired, handled, processed, and stored~\citep{ai4h2020del5-1,ai4h2020del5-4}. This documentation should include precise details of the hardware and software versions used and also cover the creation of reference annotations. Moreover, quality criteria for rejecting data and procedures for handling missing data~\citep{stevens2020} should be reported, i.e., aspects of what is \emph{not} in the dataset. Protocols should be defined prior to data acquisition when prospectively collecting test data. Completeness and clarity of the protocols should be verified during data acquisition.

Reported information should characterize the acquired dataset in a useful way. For example, summary statistics allow an initial assessment whether a given dataset is an adequate sample of the target population. Relevant subsets and biases identified in the dataset should be reported as well. Generally, one should collect and report as much information as feasible with the available resources, since retrospectively obtaining missing metadata is hard or impossible. If there will be multiple versions of a dataset, e.g., due to iterative data acquisition or review of reference annotations, versioning is needed. Suitable hashing can guarantee integrity of the entire dataset as well as its individual samples, and identify datasets without disclosing contents.

\subsection{Regulatory requirements}

AI solutions in pathology are in vitro diagnostic medical devices (IVDMDs) because they evaluate tissue images for diagnostic purposes outside the human body. Therefore, regulatory approval is required for sale and use in a clinical setting~\citep{homeyer2021}. The U.S. Food and Drug Administration (FDA) and European Union (EU) impose similar requirements to obtain regulatory approval. This includes compliance with certain quality management and documentation standards, a risk analysis, and a comprehensive performance evaluation. The performance evaluation must demonstrate that the method provides accurate and reliable results compared to a gold standard (analytical performance) and that the method provides real benefit in a clinical context (clinical performance). Good test datasets are an essential prerequisite for a meaningful evaluation of analytical performance.

\subsubsection{EU + UK}

In the EU and UK, IVDMDs are regulated by the In vitro Diagnostic Device Regulation (IVDR, formally ``Regulation 2017/746'')~\citep{eu2017}. After a transition period, compliance with the IVDR will be mandatory for novel routine pathology diagnostics as of May 26, 2022. The IVDR does not impose specific requirements on test datasets used in the analytical performance evaluation. However, the EU has put forward a proposal for an EU-wide regulation on harmonized rules for the assessment of AI~\citep{eu2021proposal}.

The EU proposal~\citep{eu2021proposal} considers AI-based IVDMDs as ``high-risk AI systems'' (preamble (30)). For test datasets used in the evaluation of such systems, the proposal imposes certain quality criteria: test datasets must be ``relevant, representative, free of errors and complete'' and ``have the appropriate statistical properties'' (Article 10.3). Likewise, it requires test datasets to be subject to ``appropriate data governance and management practices'' (preamble (44)) with regard to design choices, suitability assessment, data collection, and identification of shortcomings.

\subsubsection{USA}

In the US, IVDMDs are regulated in the Code of Federal Regulations (CFR) Part 809~\citep{ecfr21ivdr}. Just like the IVDR, the CFR does not impose specific requirements on test datasets used in the analytical performance evaluation. However, the CFR states that products should be accompanied by labeling stating specific performance characteristics (e.g., accuracy, precision, specificity, and sensitivity) related to normal and abnormal populations of biological specimens.

In 2021, the FDA approved the first AI software for pathology~\citep{fda2021prostate}. In this context, the FDA has established a definition and requirements for approval of generic AI software for pathology, formally referred to as ``software algorithm devices to assist users in digital pathology''~\citep{fda2021paigeprostateapproval}.

Test datasets used in analytical performance studies are expected to contain an ``appropriate'' number of images. To be ``representative of the entire spectrum of challenging cases'' (3.ii.A. and B. of source~\citep{fda2021paigeprostateapproval}) that can occur when the product is used as intended, test datasets should cover multiple operators, slide scanners, and clinical sites and contain ``clinical specimens with defined, clinically relevant, and challenging characteristics.''(3.ii.B. of source~\citep{fda2021paigeprostateapproval}) In particular, test datasets should be stratified into relevant subsets (e.g., by medical characteristics, patient demographics, scanning equipment) to allow separate determination of performance for each subset. Case cohorts considered in clinical performance studies (e.g., evaluating unassisted and software-assisted evaluation of pathology slides with intended users) are expected to adhere to similar specifications.

Product labeling according to CFR 809 was also defined in more detail. In addition to the general characteristics of the dataset (e.g., origin of images, annotation procedures, subsets, \ldots), limitations of the dataset (e.g., poor image quality or insufficient sampling of certain subsets) that may cause the software to fail or operate unexpectedly should be specified.

In summary, there are much more specific requirements for test datasets in the US than in the EU. However, none of the regulations clearly specify how the respective requirements can be achieved or verified.

\section{Discussion}

\begin{figure*}[t]
\centering
\includegraphics[width=.8\textwidth]{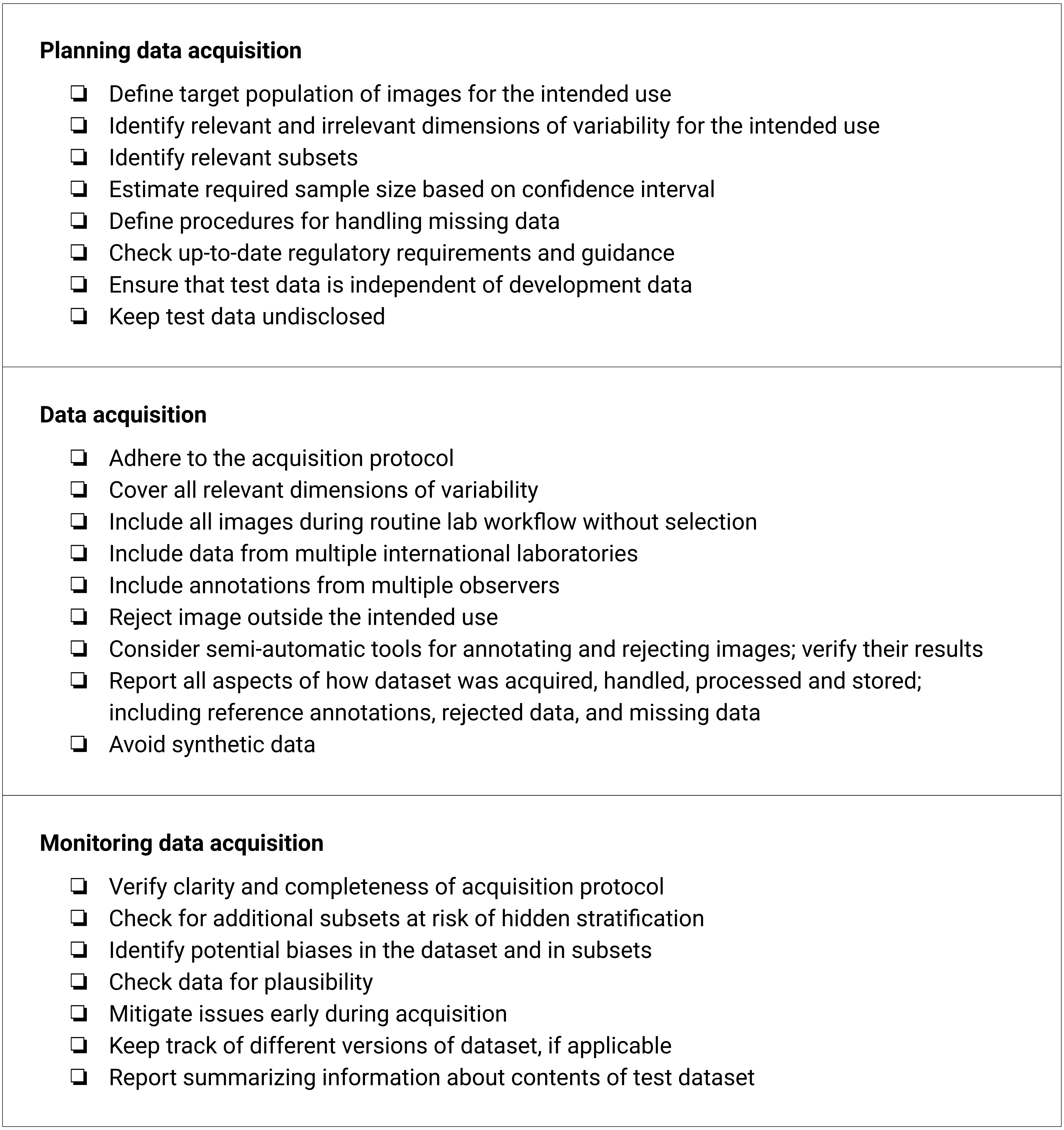}

\caption{Overview of recommendations to be considered during different phases of collecting test datasets}
\label{fig:recommendations_overview}
\end{figure*}

Our recommendations for compiling test datasets are summarized in \autoref{fig:recommendations_overview}. They are intended to help AI developers demonstrate the robustness and practicality of their solutions to regulatory agencies and end users. Likewise, the advice can be used to check whether test datasets used in the evaluation of AI solutions were appropriate and reported performance measures are meaningful. Much of the advice can also be transferred both to image analysis solutions without AI and to similar domains where solutions are applied to medical images, such as radiology or ophthalmology.

A key finding of the work is that it remains challenging to collect test datasets and that there are still many unanswered questions. The current regulatory requirements remain vague and do not specify in detail important aspects such as the required diversity of test datasets or the required confidence in measured performance metrics. The main challenge is that the target population of images is elusive, i.e., it cannot be formally specified but only roughly described. This makes it difficult to determine whether a dataset is representative, i.e., whether the many dimensions of variability are covered sufficiently, and whether the sample distribution corresponds to real-world data. Without a clear measure of representativity, it is also impossible to determine whether a test dataset is large enough to enable assessment of performance metrics with a maximum sampling error.

For regulatory approval, a plausible justification is needed why the test dataset used was good enough. Besides following the advice in this paper, it can also be helpful to refer to published studies in which AI solutions have been comprehensively evaluated. Additional guidance can be found in the summary documents of approved AI solutions published by the FDA, which include information on their evaluation~\citep{wu2021}. It turns out that many of the AI devices approved by the FDA were evaluated only at a small number of sites~\citep{wu2021} with limited geographic diversity~\citep{kaushal2020}. Test sets used in current studies typically involved 1000s of slides, 100s of patients, \textless5~sites, and \textless5~scanner types~\citep{perincheri2021,ianni2020,bulten2022,dudgeon2021}.

Today, AI solutions in pathology may not be used for primary diagnosis, but only in conjunction with a standard evaluation by the pathologist~\citep{fda2021paigeprostateapproval}. Therefore, compared to a fully automated usage scenario, requirements for robustness are considerably lower. This also applies to the expected confidence in the performance measurement and the scope of the test dataset used. In a supervised usage scenario, the accuracy of an AI solution determines how often the user needs to intervene to correct results, and thus its practical usefulness. End users are interested in the most meaningful evaluation of the accuracy of AI solutions to assess their practical utility. Therefore, a comprehensive evaluation of the real-world performance of a product, taking into account the advice given in this paper, can be an important marketing tool.

\subsection{Limitations and outlook}

Some aspects of compiling test datasets were not considered in this article. One aspect is how to collaborate with data donors, i.e., how to incentivize or compensate them for donating data. Other aspects include the choice of software tools and data formats for the collection and storage of data sets or how the use of test datasets should be regulated. These aspects must be clarified individually for each use case and the AI solution to be tested. Furthermore, we do not elaborate on legal aspects of collecting test datasets, e.g., obtaining consent from patients, privacy regulations, licensing, and liability. For more details on these topics, we refer to other works~\citep{rodrigues2020}. This paper focuses exclusively on the compilation of test datasets. For advice on other issues related to validating AI solutions in pathology, such as how to select an appropriate performance metric, how to make algorithmic results interpretable, or how to conduct a clinical performance evaluation with end users, we also refer to other works~\citep{maleki2020,kelly2019,oala2020,park2021,dehond2022,evans2022}.

For AI solutions to operate with less user intervention and to better support diagnostic workflows, real-world performance must be assessed more accurately than is currently possible. The key to accurate performance measures is the representativeness of the test dataset. Therefore, future work should focus on better characterizing the target population of images and how to collect more representative samples. Empirical studies should be conducted on how different levels of coverage of the variability dimensions (e.g., laboratories, scanner types) affect the quality of performance evaluation for common use cases in computational pathology.

In addition, clear criteria should be developed to delineate the target population from unsuitable data. Currently, the assessment of the suitability of data is typically done by humans, which might introduce subjective bias. Automated methods can help to make the assessment of suitability more objective (see ``Curation'') and should therefore be further explored. However, such automated methods must be validated on dedicated test datasets themselves.

Another open challenge is how to deal with changes in the target population of images. Since the intended use for a particular product is fixed, in theory the requirements for the test datasets should also be fixed. However, the target distribution of images is influenced by several factors that change over time. These include technological advances in specimen and image acquisition, distribution of scanner systems used, and shifting patient populations~\citep{finlayson2021,kelly2019}. As part of post-market surveillance, AI solutions must be continuously monitored during their entire lifecycle~\citep{MDCG2022-2}. Clear processes are required for identifying changes in the target population of images and adapting performance estimates accordingly.

\section{Conclusions}

Appropriate test datasets are essential for meaningful evaluation of the performance of AI solutions. The recommendations provided in this article are intended to help demonstrate the utility of AI solutions in pathology and to assess the validity of performance studies. The key remaining challenge is the vast variability of images in computational pathology. Further research is needed on how to formalize criteria for sufficiently representative test datasets so that AI solutions can operate with less user intervention and better support diagnostic workflows in the future.

\section*{Acknowledgments}

A.H., C.G., L.O.S., F.Z., T.E., K.S., A.K., C.O.R., T.S. R.C., P.B., P.H., and N.Z. were supported by the German Federal Ministry for Economic Affairs and Climate Action via the EMPAIA project (grant numbers 01MK20002A, 01MK20002B, 01MK20002C, 01MK20002E). M.Ka., M.P., and H.M. received funding from the Austrian Science Fund (FWF), Project P-32554 (Explainable Artificial Intelligence), the Austrian Research Promotion Agency (FFG) under grant agreement No.~879881 (EMPAIA), and the European Union's Horizon 2020 research and innovation programme under grant agreement No.~857122 (CY-Biobank). P.S. was funded by Helmholtz Association's Initiative and Networking Fund through Helmholtz AI. R.B. was supported by the START Program of the Faculty of Medicine of the RWTH Aachen University (Grant-Nr. 148/21). P.B. was also supported by the German Research Foundation (DFG, Project IDs 322900939, 454024652, 432698239, and 445703531), the European Research Council (ERC, CoG AIM.imaging.CKD No.~101001791), the German Federal Ministries of Health (Deep Liver, No.~ZMVI1-2520DAT111), Education and Research (STOP-FSGS-01GM1901A). The funders had no role in the committee work, discussions, literature research, decision to publish, or preparation of the manuscript.

\section*{Author Contributions}

A.H. and C.G. organized the committee work. A.H., C.G., and L.O.S. conceived the manuscript. A.H., C.G., L.O.S., F.Z., T.E., K.S., M.W., and R.D.B. wrote the manuscript. All authors participated in the committee work and contributed to the literature review. The final version of the paper was reviewed and approved by all authors.

\section*{Competing interests}

F.Z. is a shareholder of asgen GmbH. P.S. is a member of the supervisory board of asgen GmbH. All other authors declare that they have no conflict of interest.

\bibliography{ms}

\end{document}